# Detuning dependence of high-order harmonic generation in monolayer transition metal dichalcogenides

Tomohiro Tamaya[1,2][*], Satoru Konabe[3], and Shiro Kawabata[1,2]

[1]*Nanolectronics Research Institute(NeRI), National Institute of Advanced Industrial Science and Technology(AIST), Tsukuba, Ibaraki 305-8568, Japan*
[2]*CREST, Japan Science and Technology Agency, Kawaguchi, Saitama 332-0012, Japan*
[3]*Research Institute for Science and Technology, Tokyo University of Science, Katsushika, Tokyo 125-8585, Japan*

[*]E-mail: t-tamaya@aist.go.jp

We theoretically investigate detuning-dependent properties of high-order harmonic generation (HHG) in monolayer transition metal dichalcogenides (TMDCs). In contrast to HHG in conventional materials, TMDCs show both parallel and perpendicular emissions with respect to the incident electric field. We find that such an anomalous emission can be artificially controlled by the frequency detuning of the incident electric fields, i.e., the parallel and perpendicular HHG can be strongly enhanced by multiphoton resonances. This peculiar phenomenon would provide a way for controlling HHG in TMDCs and stimulate the realization of novel optical devices.

# 1. Introduction

Atomically thin two-dimensional materials have been actively investigated in recent years because of their potential applications for electronics and optics. Among them, the transition metal dichalcogenides (TMDCs) are one of the representatives [1-3] having a two-dimensional hexagonal lattice of M (M = Mo or W) and X (X = S or Se) atoms with the inversion symmetry breaking. The inversion symmetry breaking of the crystal structure gives rise to the energy band gap at the $K^{\pm}$ points, which provides a fascinating platform for valley physics [4–8], giving unique optical properties such as the valley dependent optical selection rules for interband transitions [9-12].

On the other hand, high-order harmonic generation (HHG) in solid-state materials caused by intense light has attracted much attentions in recent years [13-18]. The HHG in TMDCs has also been intensively studied experimentally [19-24], because of the expectation to show atypical light-matter interactions in contrast to the conventional solid-state materials. A recent experimental study found that the TMDCs have a peculiar characteristic of HHG [23], i.e., show both parallel and perpendicular HHG emissions with respect to the incident electric fields (Fig. 1). More peculiarly, the parallel (perpendicular) emissions mainly contain only odd-(even-) order harmonics. Recently, we have shown theoretically that these anomalous behaviors can be explained by the wavenumber dependences of the dipole moments, which are fundamentally determined by the symmetries of both atomic orbitals and lattice structures [25]. Thus, the TMDCs are identified to show unique characteristics of HHG and development of the control method for HHG would be demanded for opening new optical applications of TMDCs.

In this paper, we will theoretically propose a way to artificially control both parallel and perpendicular emissions of HHG in TMDCs, by tuning the frequency of the incident electric fields. We have shown the $n$th-ordered harmonics can be enhanced when the frequency of the incident electric fields is set to one $n$th. This peculiar phenomenon would provide a way to control the both perpendicular and parallel emissions of HHG and stimulate the realization of novel optical devices.

## 2. Theory

The theoretical framework utilized in this paper was constructed by applying the theory we have established previously [26, 27] to the case of TMDCs. The framework is based on the tight-binding model, where the honeycomb lattice $MX_2$ is constructed with M (S, Se, Te) and X (Mo, W) atoms. In this model, we will assume the wave-functions of M and X atoms as $\phi_A(x) = p_x(x) + ip_y(x)$ and $\phi_B(x) = d_{xy}(x) + i\tau^z d_{x^2-y^2}(x)$, respectively. Here, $\tau^z = \pm 1$ is the variable describing the state at the $K^\pm$ points. In addition, we will define the difference of the onsite energy for each atom as $m$. The Rabi frequency in this model can be represented as $\Omega_R(\mathbf{k}, t) = (e/m_0 c)\Sigma_i e^{i\mathbf{k}\cdot\boldsymbol{\delta}_i} \int d^2x\, \phi_B^*(x) \mathbf{A}(t)\cdot \mathbf{p}\phi_A(\mathbf{x} - \boldsymbol{\delta}_i)$ [26, 27]. Here, $\boldsymbol{\delta}_i$ is the lattice vector, $m_0$ is the electron mass, $e$ is the electron charge, $c$ is the velocity of light, $\mathbf{k}$ is the Bloch wavevector, $\mathbf{p}$ is the momentum of the electron, and $\mathbf{A}(t)$ is the vector potential defined as $\mathbf{A}(t) = A_0(\cos\omega_0 t, 0)$ where $\omega_0$ is the frequency of the incident electric field. In this case, we can express the Rabi frequency near the $K^\pm$ points as $\Omega_R(\mathbf{k}, t) \approx \Omega_{R0}(t)[[\tau^z + \beta k_x a]\cos\omega_0 t + i\tau^z \gamma k_y a]$, where $\beta = -0.27$ and $\gamma = 0.87$ are the dimensionless constants, $\Omega_{R0}(t) = \Omega_{R0} \exp[-(t - t_0)^2/T^2]$, and $a$ is the lattice constant [25]. Here, we define the time-independent Rabi frequency as $\Omega_{R0}$. Throughout this paper, we set $t_0 = 12\pi/\omega_0$ and $T = 4\pi/\omega_0$, respectively.

Considering the above expressions and performing the same procedure as in Refs [26, 27], we can derive the time evolution equations of the carrier densities $f_k^{\sigma,\tau^z} = \langle \sigma_k^\dagger \sigma_k \rangle (\sigma = e, h)$ and the polarization $P_k^{\tau^z} = \langle h_{-k}^\dagger e_k \rangle$ for $\tau^z = \pm 1$ in the form [25],

$$
\begin{aligned}
i\frac{\partial}{\partial t} P_k^{\tau^z} =\ & 2[\epsilon_k^e(t) + \epsilon_k^h(t)]P_k^{\tau^z} \\
& +\hbar(m/E_k)(\tau^z \cos\theta_k \operatorname{Re}[\Omega_R(\mathbf{k},t)] - \sin\theta_k \operatorname{Im}[\Omega_R(\mathbf{k},t)])\left[1 - f_k^{e,\tau^z} - f_k^{h,\tau^z}\right]P_k^{\tau^z} \\
& +i\hbar(\sin\theta_k \operatorname{Re}[\Omega_R(\mathbf{k},t)] + \tau^z \cos\theta_k \operatorname{Im}[\Omega_R(\mathbf{k},t)])\left[1 - f_k^{e,\tau^z} - f_k^{h,\tau^z}\right] - i\gamma_t P_k^{\tau^z},
\end{aligned}
\quad (1)
$$

$$
\begin{aligned}
\frac{\partial}{\partial t} f_k^{\sigma,\tau^z} =\ & -2\tau^z\left[(\tau^z \sin\theta_k \operatorname{Re}[\Omega_R(\mathbf{k},t)] + \cos\theta_k \operatorname{Im}[\Omega_R(\mathbf{k},t)])\operatorname{Im}[(iP_k^{\tau^z})^\dagger]\right. \\
& \left. + 2(m/E_k)(\tau^z \cos\theta_k \operatorname{Re}[\Omega_R(\mathbf{k},t)] - \sin\theta_k \operatorname{Im}[\Omega_R(\mathbf{k},t)])\operatorname{Im}[(iP_k^{\tau^z})^\dagger]\right] - \gamma_l f_k^{\sigma,\tau^z}.
\end{aligned}
\quad (2)
$$

Here, we define $E_k^\sigma$ and $\epsilon_k^\sigma(t)$ as $E_k^\sigma = \sqrt{(\hbar v_F k)^2 + m^2}$ and $\epsilon_k^\sigma(t) = E_k + \hbar(\tau^z \sin\theta_k \text{Re}[\Omega_R(\boldsymbol{k},t)] + \cos\theta_k \text{Im}[\Omega_R(\boldsymbol{k},t)])$, respectively, where $\hbar$ and $v_F$ are the reduced Planck constant and Fermi velocity. The factors $\gamma_t$ and $\gamma_l$ are the transverse and longitudinal relaxation constants, and they are fixed to $\gamma_t = 0.1\omega_0$ and $\gamma_l = 0.01\omega_0$ [26]. The numerical solutions of these equations give the time evolutions of the distributions of the carrier densities $f_k^{\sigma,\tau^z}$ and polarization $P_k^{\tau^z}$ in two-dimensional $\boldsymbol{k}$ space. Utilizing these distributions, the time evolution of the generated currents along the *x*- and *y*- axes can be calculated using $J_\nu(t) = -c\langle \partial H_I/\partial A_\nu \rangle (\nu = x, y)$. Thus, we can obtain the time evolutions of the current in the form, $J(t) = \big(J_x(t), J_y(t)\big)$, and can calculate HHG spectra from $I(\omega) = \omega^2 |J(\omega)|^2$, where $J(\omega) = \big(J_x(\omega), J_y(\omega)\big)$ is the Fourier transform of $J(t)$.

## 3. Results and discussion

Our numerical calculation shows that both parallel and perpendicular emissions exist with respect to the incident electric fields. The parallel emissions of HHG mainly include only odd-ordered harmonics while the perpendicular emissions include the even-ordered ones. These tendencies can be attributed to the unique wave number dependence of the Rabi frequency, which is fundamentally determined by the lattice structures and wave-functions of each atom [25]. We plotted in the bottom of Fig. 2 the detuning-dependent properties of the parallel emissions of HHG for the third- (blue line), fifth- (orange line), and seventh-ordered (green line) harmonics. In our numerical calculation, we fix the Rabi frequency as $\Omega_{R0} = 0.1\omega_0$ and vary the ratio of band-gap energy and the frequency of the incident electric field, i.e., $E_g/\hbar\omega_0$. This figure clearly indicates that the third, fifth, and seventh-ordered harmonics have maximum peaks at $E_g/\hbar\omega_0 = 3, 5,$ and $7$, respectively. These behaviors can be described as resonant three-, five-, and seven-photon absorptions in the upper panels in Fig. 2. We also plotted in the bottom panel of Fig. 3 the detuning-dependent properties of the perpendicular emissions of HHG for the second- (blue line), fourth- (orange line), and sixth-ordered (green line) harmonics. Similar to the parallel emissions, the second, fourth, and sixth-ordered harmonics have maximum peaks at $E_g/\hbar\omega_0 = 2, 4$, and $6$. These conditions correspond to resonant two-, four-, and six-photon absorptions as shown in the upper panels in Fig. 3. Therefore, we can conclude that the HHG can be maximized at the resonant frequencies of the incident electric field with the band-gap energy of TMDCs.

The above results and discussions can be summarized as follows: when the frequencies of the incident electric fields are detuned at the band-gap energy, for example, $E_g/\hbar\omega_0 = 5$ (Fig. 4), then the fifth-ordered harmonics in the parallel direction and fourth- and sixth-ordered harmonics in the perpendicular direction are simultaneously enhanced. This consideration generally indicates that the detuned frequencies of the incident electric field, i.e., $E_g/\hbar\omega_0 = n$, can enhance the specific-ordered harmonics, $n$th and $(n \pm 1)$th HHG, in both parallel and perpendicular directions. This phenomenon is expected to provide a control method of HHG in TMDCs and stimulate the realization of novel optical devices, that might be useful for the self-referencing technique of optical frequency comb technologies. The usual optical frequency comb technique employs f-to-2f self-referencing scheme [28, 29], where one-octave supercontinuum (SC) spectrum width is required to interfere the fundamental and the second harmonic generations. Similar to this technique, 2f-to-3f self-referencing method [30, 31] with Mach-Zehnder interferometer has also been realized in recent years, where the SC spectrum is required to spread only for $(1/2)\omega_0$ width. This progress enables us to expect the *n*f-to-*(n+1)*f self-referencing technique in TMDCs using parallel and perpendicular emissions of HHG. This method requires the SC spectrum width only for $(1/n)\omega_0$. In this scheme, the parallel and perpendicular HHG have phase-locked property and can make compact the experimental system. Thus, HHG in TMDCs are expected to have a possibility for developing the useful method of new optical frequency comb method.

## 4. Conclusions

In conclusion, we have theoretically investigated the detuning dependence of HHG in monolayer TMDCs. TMDCs show both parallel and perpendicular emissions with respect to the incident electric fields. By our numerical calculations, we found that such an anomalous emission can be maximized by the multiphoton resonance conditions. Our results would provide a control method of HHG in the parallel and perpendicular directions and stimulate the realization of novel optical devices in optical frequency comb method.

## Acknowledgments

This work was supported by JST CREST (JPMJCR14F1), JST Nanotech CUPAL, and MEXT KAKENHI(15H03525).

## Figure Captions

**Fig. 1.** (Color online) Schematic diagrams of HHG in (a) real-space and (b) *k*-space pictures. These figures indicate that the both parallel and perpendicular HHG with respect to the incident electric fields exist in TMDCs.

**Fig. 2.** (Color) (Bottom figure): Detuning-dependent properties of HHG in the parallel emissions with respect to the incident electric fields. Blue, orange, and green lines indicate the third, fifth, and seventh-order harmonics in TMDCs. (Upper figure): Schematic diagrams of three-, fifth-, and seventh-multiphoton resonances in TMDCs.

**Fig. 3.** (Color) (Bottom figure): Detuning-dependent properties of HHG in the perpendicular emissions with respect to the incident electric fields. Blue, orange, and green lines indicate the second, forth, and sixth-order harmonics in TMDCs. (Upper figure): Schematic diagrams of two-, fourth-, and sixth-multiphoton resonances in TMDCs.

**Fig. 4.** (Color) Schematic diagram of the incident light and HHG in TMDCs for the case of $E_g/\hbar\omega_0 = 5$. In this case, the fifth-order harmonics in the parallel direction and fourth- and sixth-order emissions in the perpendicular direction are emphasized.

**Fig.1. (Color Online)**

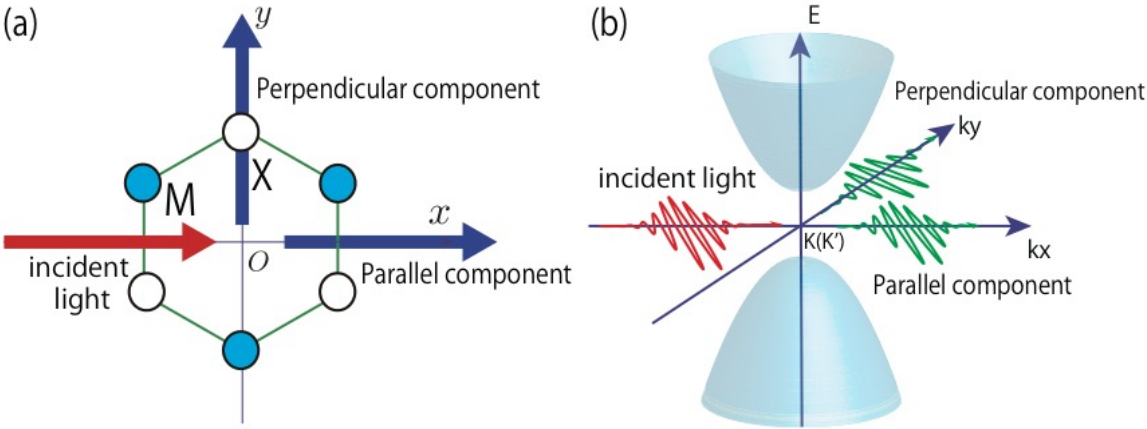

**T. Tamaya** *et al*.

**Fig.2.** (Color Online)

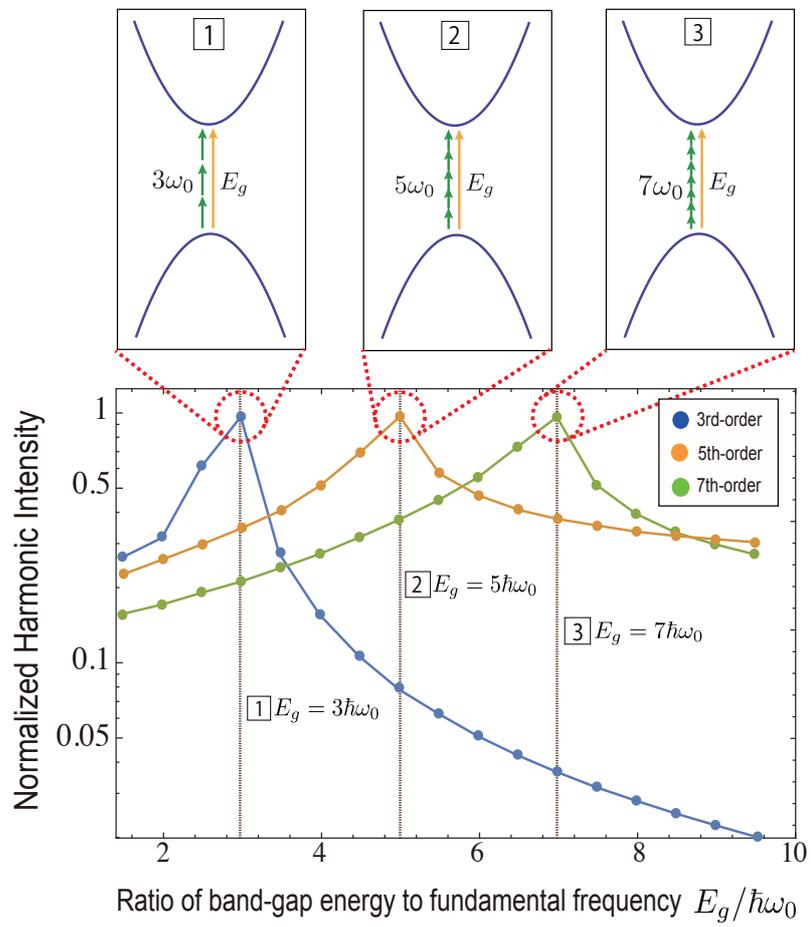

**T. Tamaya** *et al.*

**Fig.3. (Color Online)**

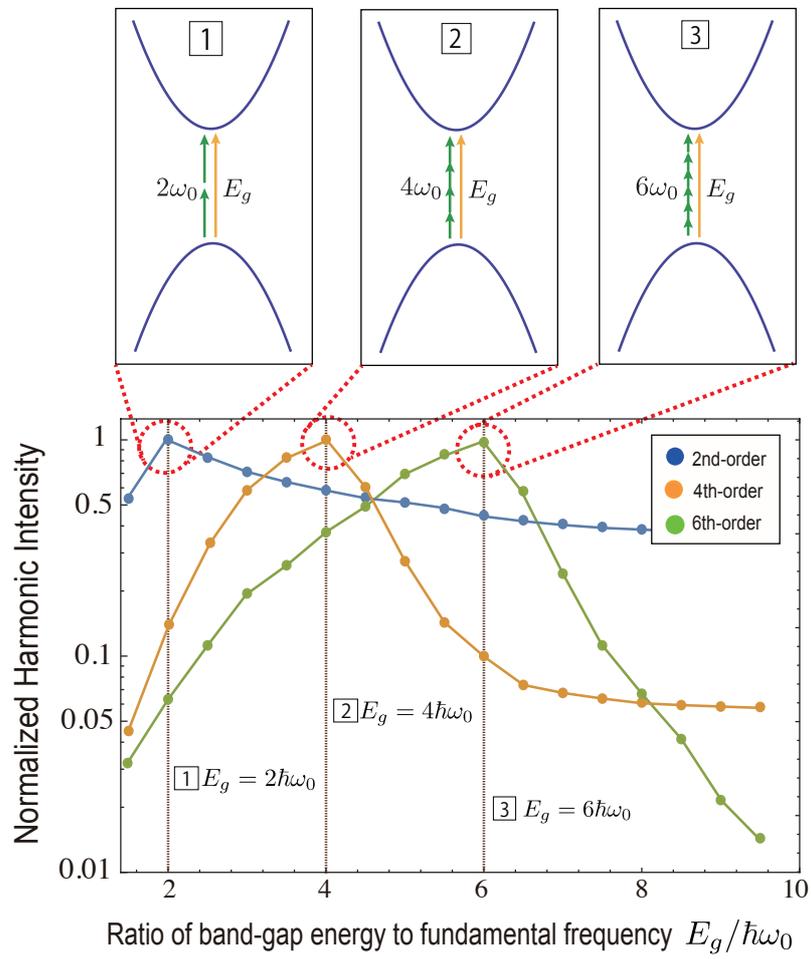
**T. Tamaya *et al*.**

**Fig.3. (Color Online)**

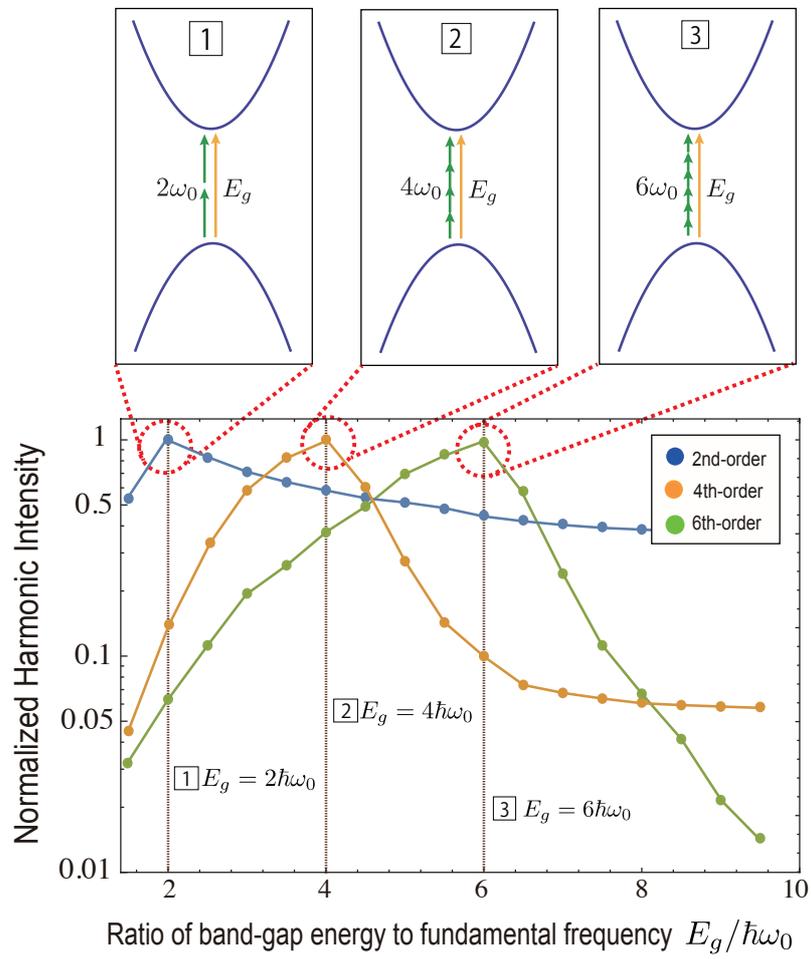

**Fig.4.  (Color Online)**

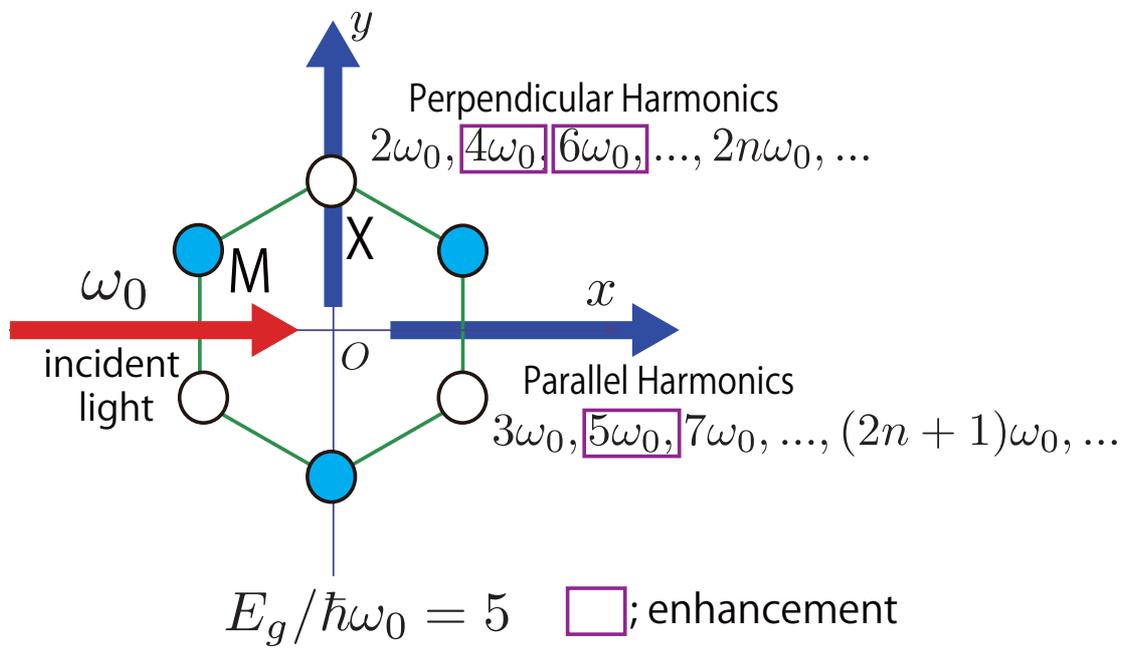

$E_g/\hbar\omega_0 = 5$ ☐; enhancement

**T. Tamaya *et al*.**